# Study of the loss of Xenon Scintillation in Xenon-Trimethylamine Mixtures

A.M.F. Trindade,[a,b] J. Escada,[a,b] A.F.V. Cortez,[a,b] F.I.G.M. Borges,[a,b] F.P. Santos,[a,b] C. Adams,[c] V. Álvarez,[d] L. Arazi,[e] C.D.R. Azevedo,[f] F. Ballester,[i] J.M. Benlloch-Rodríguez,[d] A. Botas,[d] S. Cárcel,[d] J.V. Carríon,[d] S. Cebrián,[g] C.A.N. Conde,[a,b] J. Díaz,[d] M. Diesburg,[h] R. Esteve,[i] R. Felkai,[d] L.M.P. Fernandes,[j] P. Ferrario,[d,k] A.L. Ferreira,[f] E.D.C. Freitas,[j] A. Goldschmidt,[l] J.J. Gómez-Cadenas,[d,1] D. González-Díaz,[m] R. Guenette,[c] R.M. Gutiérrez,[n] K. Hafidi,[o] J. Hauptman,[p] C.A.O. Henriques,[j] A.I. Hernandez,[n] J.A. Hernando Morata,[m] V. Herrero,[i] S. Johnston,[o] B.J.P. Jones,[q] L. Labarga,[r] A. Laing,[d] P. Lebrun,[h] I. Liubarsky,[d] N. López-March,[d] M. Losada,[n] J. Martín-Albo,[c] G. Martínez-Lema,[m] A. Martínez,[d] A.D. McDonald,[q] F. Monrabal,[q] C.M.B. Monteiro,[j] F.J. Mora,[i] L.M. Moutinho,[f] J. Muñoz Vidal,[d] M. Musti,[d] M. Nebot-Guinot,[d] P. Novella,[d] D.R. Nygren[q,1] B. Palmeiro,[d] A. Para,[h] J. Pérez,[d] M. Querol,[d] J. Renner,[d] J. Repond,[o] S. Riordan,[o] L. Ripoll,[s] J. Rodríguez,[d] L. Rogers,[q] J.M.F. dos Santos,[j] A. Simón,[d] C. Sofka,[t,2] M. Sorel,[d] T. Stiegler,[t] J.F. Toledo,[i] J. Torrent,[d] Z. Tsamalaidze,[u] J.F.C.A. Veloso,[f] R. Webb,[t] J.T. White,[t,3] N. Yahlali[d]

[a] *LIP-Laboratório de Instrumentação e Física Experimental de Partículas, Coimbra, Portugal*
[b] *Departamento de Física da Universidade de Coimbra, Rua Larga 3004-516 Coimbra, Portugal*
[c] *Department of Physics, Harvard University, Cambridge, MA 02138, USA*
[d] *Instituto de Física Corpuscular (IFIC), CSIC & Universitat de València, Calle Catedrático José Beltrán, 2, 46980 Paterna, Valencia, Spain*
[e] *Nuclear Engineering Unit, Faculty of Engineering Sciences, Ben-Gurion University of the Negev, P.O.B. 653 Beer-Sheva 8410501, Israel*

---

1 NEXT Co-spokesperson.
2 Now at University of Texas at Austin, USA.
3 Deceased.

[f] *Institute of Nanostructures, Nanomodelling and Nanofabrication (i3N), Universidade de Aveiro, Campus de Santiago, 3810-193 Aveiro, Portugal*
[g] *Laboratorio de Física Nuclear y Astropartículas, Universidad de Zaragoza, Calle Pedro Cerbuna, 12, 50009 Zaragoza, Spain*
[h] *Fermi National Accelerator Laboratory, Batavia, Illinois 60510, USA*
[i] *Instituto de Instrumentación para Imagen Molecular (I3M), Centro Mixto CSIC - Universitat Politècnica de València Camino de Vera, s/n, 46022 Valencia, Spain*
[j] *LIBPhys, Physics Department, University of Coimbra, Rua Larga, 3004-516 Coimbra, Portugal*
[k] *Donostia International Physics Center (DIPC), Paseo Manuel Lardizabal 4, 20018 Donostia-San Sebastian, Spain*
[l] *Lawrence Berkeley National Laboratory (LBNL), 1 Cyclotron Road, Berkeley, California 94720, USA*
[m] *Instituto Gallego de Física de Altas Energías, Univ. de Santiago de Compostela, Campus sur, Rúa Xosé María Suárez Núñez, s/n, 15782 Santiago de Compostela, Spain*
[n] *Centro de Investigación en Ciencias Básicas y Aplicadas, Universidad Antonio Nariño, Sede Circunvalar, Carretera 3 Este No. 47 A-15, Bogotá, Colombia*
[o] *Argonne National Laboratory, Argonne IL 60439, USA*
[p] *Department of Physics and Astronomy, Iowa State University 12, Physics Hall, Ames, Iowa 50011-3160, USA*
[q] *Department of Physics, University of Texas at Arlington, Arlington, Texas 76019, USA*
[r] *Departamento de Física Teórica, Universidad Autónoma de Madrid, Campus de Cantoblanco, 28049 Madrid, Spain*
[s] *Escola Politècnica Superior, Universitat de Girona, Av. Montilivi, s/n, 17071 Girona, Spain*
[t] *Department of Physics and Astronomy, Texas A&M University, College Station, Texas 77843-4242, USA*
[u] *Joint Institute for Nuclear Research (JINR), Joliot-Curie 6, 141980 Dubna, Russia*

**Abstract**

This work investigates the capability of TMA ($(CH_3)_3N$) molecules to shift the wavelength of Xe VUV emission (160-188 nm) to a longer, more manageable, wavelength (260-350 nm). Light emitted from a Xe lamp was passed through a gas chamber filled with Xe-TMA mixtures at 800 Torr and detected with a photomultiplier tube. Using bandpass filters in the proper transmission ranges, no reemitted light was observed experimentally. Considering the detection limit of the experimental system, if reemission by TMA molecules occurs, it is below 0.3% of the scintillation absorbed in the 160-188 nm range. An absorption coefficient value for xenon VUV light by TMA of 0.43±0.03 $cm^{-1}Torr^{-1}$ was also obtained. These results can be especially important for experiments considering TMA as a molecular additive to Xe in large volume optical time projection chambers.

## 1. Introduction

High-pressure gaseous xenon optical time projection chambers (OTPCs) hold high promise for neutrinoless double beta decay ($\mathbf{0\nu\beta\beta}$) searches in $^{136}$Xe, with strong background rejection based on high energy resolution and accurate track reconstruction. Presently, two main collaborations are developing the technology towards ton-scale masses: NEXT [1-3] and more recently PandaX-III [4], with early R&D also carried out by the AXEL collaboration [5]. The baseline design of NEXT consists of pure Xe with no additives. This allows aiming at a superb energy resolution of 0.5% FWHM at the Q-value of the decay ($\boldsymbol{Q_{\beta\beta} = 2458\ \mathrm{keV}}$) relying on electroluminescence (EL) with no charge multiplication. [6-8]. However, elastic collisions of the drifting electrons with the heavy Xe atoms result in considerable diffusion which degrades the quality of track imaging. Electron diffusion can be significantly reduced by introducing low concentrations of molecular additives to Xe [9, 10]. These cool down the electrons by including inelastic collisions that transfer kinetic energy to internal degrees of freedom, but at the price of degraded energy resolution. This compromise calls for detailed experimental studies to assess the positive and negative effect of such doping.
One of the molecular additives suggested [7] for high-pressure Xe OTPCs for $\mathbf{0\nu\beta\beta}$ searches is trimethylamine (TMA, $(CH_3)_3N$). TMA is potentially a promising option, since it might have the additional advantage of shifting the wavelength of vacuum ultraviolet xenon scintillation [11] centred at 172 nm, to a higher, more manageable wavelength**,** eventually avoiding the use of deposited wavelength converters [12] that can also present some problems, namely in the gas purity.

Furthermore, the idea in [7] was to convert Xe excitation at the primary track to TMA ionization by the Penning effect [13, 14], thereby reducing the Fano factor and improving the energy resolution. These advantages may eventually compensate for the decrease in scintillation yield that is usually associated to the presence of molecular additives [10, 15, 16].

The NEXT TPC uses primary Xe scintillation emitted at the ionization track to determine the start time $t_0$ of the event, allowing to calculate the longitudinal coordinate of the event.

Even though this signal is strongly suppressed by small amounts of TMA [17, 18], the large number of emitted photons may, in principle, still be enough for a robust $t_0$ determination. Figure 1 shows a map of the relevant processes affecting this question. This includes Penning transfers between excited Xe states and TMA, and charge transfers between positive Xe ions and TMA - both of which may, in principle, lead to light emission by TMA during recombination or de-excitation; in addition, the map includes the absorption of Xe VUV photons and possible reemission by TMA at a longer wavelength, which is the subject of the present study (dashed rectangle). Previous works have shown that TMA absorbs light (in the range 115-260 nm) [19-23]. However, reemission studies, only carried out from 210 to 260 nm, concluded that reemission occurs, partially or totally, depending on the absorbed wavelength, in the 260-350 nm range [22, 23]. Our purpose here is to extend these studies to the absorption of Xe VUV light, noting that emission at longer wavelengths is favorable with respect to the photon detection efficiency of silicon photomultipliers. As a useful by-product, we also re-measure the absorption coefficient of Xe VUV light in Xe-TMA.

To make this study, an experimental setup was devised, which will be described and discussed in the following section.

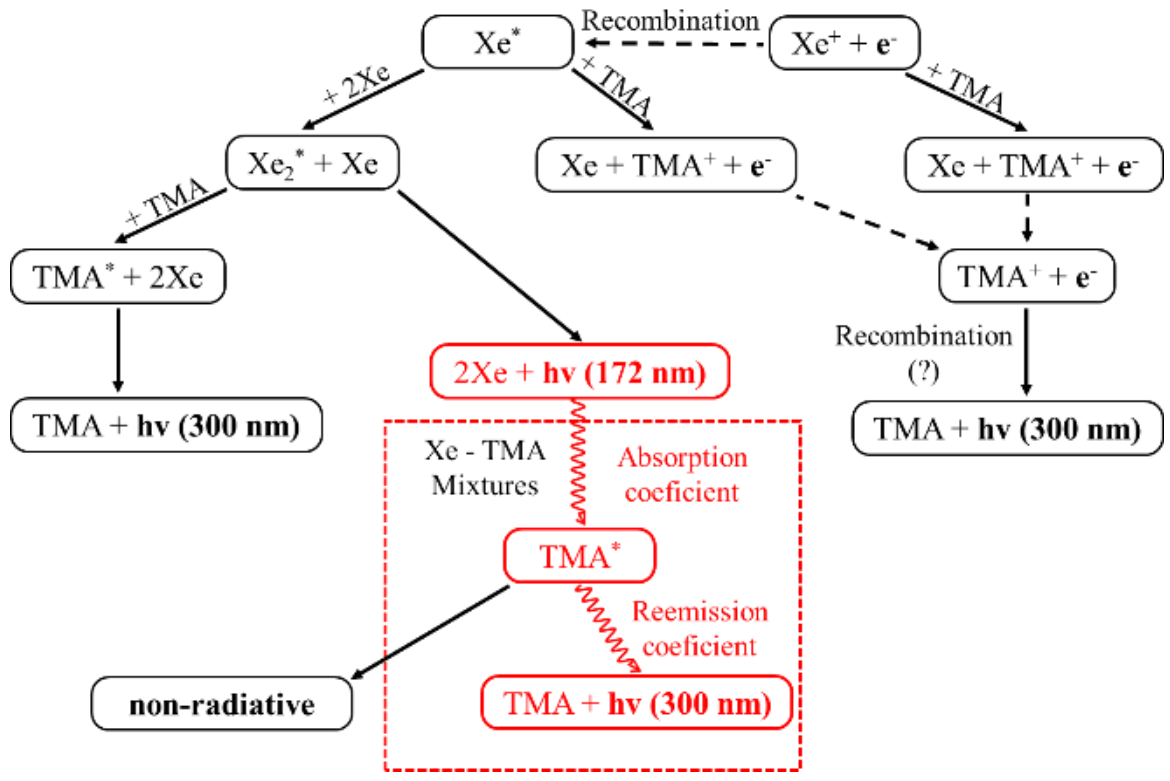

Fig. 1. Schematic of the reactions in Xe-TMA mixtures after ionization and excitation of Xe. The present work studied the processes in red.

## 2. Experimental setup and method

The experimental setup consisted of a cylindrical stainless steel chamber with two opposed apertures (up and bottom bases) and two connections to the gas system, as can be seen in Fig. 2.

The chamber was 49.7 mm long with its upper base in contact with a xenon lamp through a suprasil® window (311 suprasil® Heareus, 10 mm thick) and the bottom base connected to a photomultiplier tube (Hamamatsu model R8520-406), through a second similar window. These windows and the photomultiplier tube (PMT) are suitable for the transmission and detection of light in the wavelength range of interest. The window has an increasing transmission efficiency from about 5% at 160 nm to 85% at 180 nm and to 92% at 500 nm [24]. The PMT has a quantum efficiency slightly above 20% for the wavelengths of interest [25]. When needed, adequate bandpass filters were fitted between the PMT and the window, as shown in Fig. 2, carefully adjusting these three surfaces, to minimize the air absorption for these wavelengths.

The xenon lamp used was custom made (it is a xenon-filled proportional counter with an $^{241}$Am radioactive source placed inside), with the light intensity controlled by the voltage applied to the anode. Since in a proportional counter without gas purification the scintillation intensity decreases with time due to the increase of the impurity levels in the gas filling, the light intensity emitted by this lamp was monitored along the experiment and appropriate corrections were made for this effect.

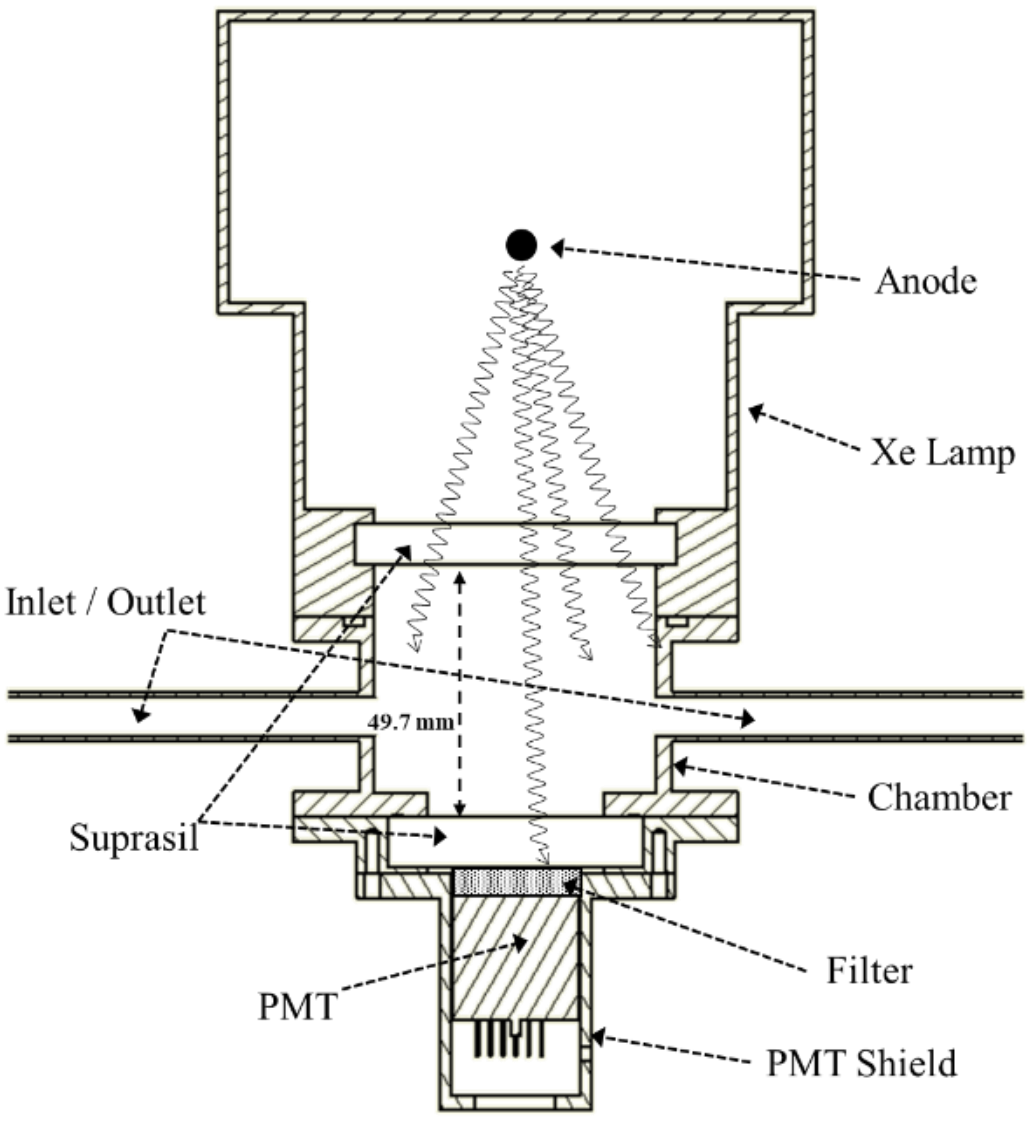

Fig. 2. Schematic of the experimental setup, showing the custom made Xe lamp, separated from the gas chamber through a suprasil® window. Also shown is the photomultiplier tube (PMT) that collects the light from the gas chamber and that is also connected to the gas chamber through another suprasil® window and eventually a filter.

For each measurement, the chamber was filled with the gas to be studied and, when the xenon lamp was turned on, xenon scintillation photons entered the gas filled chamber. If absorbed by the gas molecules, these photons may be reemitted at longer wavelengths. The light collected at the PMT, corresponding to light that was either not absorbed in the gas or absorbed and reemitted, produced a signal which was fed to a multichannel analyzer (MCA - Amptek MCA8000D), generating a spectrum with a centroid proportional to the number of photons collected per lamp event.

In order to single out the range of wavelengths of interest for this study – Xe scintillation emission (160-188 nm), coming directly from the lamp, and TMA reemission (260-350 nm) – appropriate bandpass filters were used, whose characteristics are summarized in table 1. As shown, the filters had different nominal transmissions for which the corresponding peak centroid positions in the MCA were corrected. Filter BP1 (U-330 UV from Edmund Optics) had a transmission above 70% in the 250-370 nm range with the maximum of 90% at 310 nm [26]. Filter BP2 (VUV bandpass from S.A.Matra®) had a peak nominal transmission at 172 nm of 12% with 17 nm FWHM [27]. Nevertheless, since the range of wavelengths transmitted by this filter (160-188 nm) corresponds to a region of abrupt increase of the suprasil® window transmission [24], this transmission was corrected by weighting the bandpass filter transmission with that of the suprasil® window in the range of wavelengths transmitted by the filter. Thus, an average transmission of 9.5% was obtained for the bandpass filter. Without filter (WF) the transmission range (160-650 nm) was limited by the PMT response and window transmission.

Table 1

Nominal transmission efficiency for the filters used

| Filter type | Transm. range (nm) | Max.transm. eff. (%) |
|---|---|---|
| Without filter (WF) | 160-650 | 100 |
| Bandpass 1 (BP1) | 220-400 | 90 |
| Thin glass (TG) | >350 | 95 |
| Bandpass 2 (BP2) | 160-188 | 12 |

The full lamp emission spectrum was not available, although it was known to be predominantly in the Xe VUV range. To have a better understanding of the experimental results, a study of the lamp emission using the different filters was made, with vacuum in the experimental chamber ($10^{-5}$ Torr). Fig. 3 shows the light collected when using each of the filters, corrected by the respective transmission of the filter used, and normalized to the total light collected without filter. It can be confirmed that the lamp emitted mainly (>90%) in the 160-188 nm range, however residual emission (<0.75%) above 220 nm was also detected. Although the set of filters used covered the regions of interest for the present study (160-188 and 260-350 nm), allowing to separate the relevant ranges, the overall filter coverage has a gap in the 188-220 nm range which means that photons within the 188-220 nm range were only detected without filter. As can be seen in Fig. 3, the sum of the contributions of the different wavelength ranges amounts to 94%. The missing 6% can be due to either the photons emitted in the filter gap range (188-220 nm), or to a slight overestimation of the BP2 filter efficiency. Nevertheless, since the reported TMA reemission is in the 260-350 nm range, this limitation was not considered relevant for the present study.

Prior to each gas filling, high vacuum was made in the chamber and during the measurements, the gas composition of each mixture was monitored with a residual gas analyzer - RGA (Hiden quadrupole HAL200), placed on the evacuation line, isolated by a precision leak valve, through which a small leak was allowed, enabling the gas analysis. The RGA was previously calibrated for these mixtures.

The working gas pressure was 800 Torr for all the experiments and the gas was continuously purified by convection, either with a hot getter SAES 707 in pure xenon or a cold getter SAES MC1-702-F in Xe-TMA mixtures.

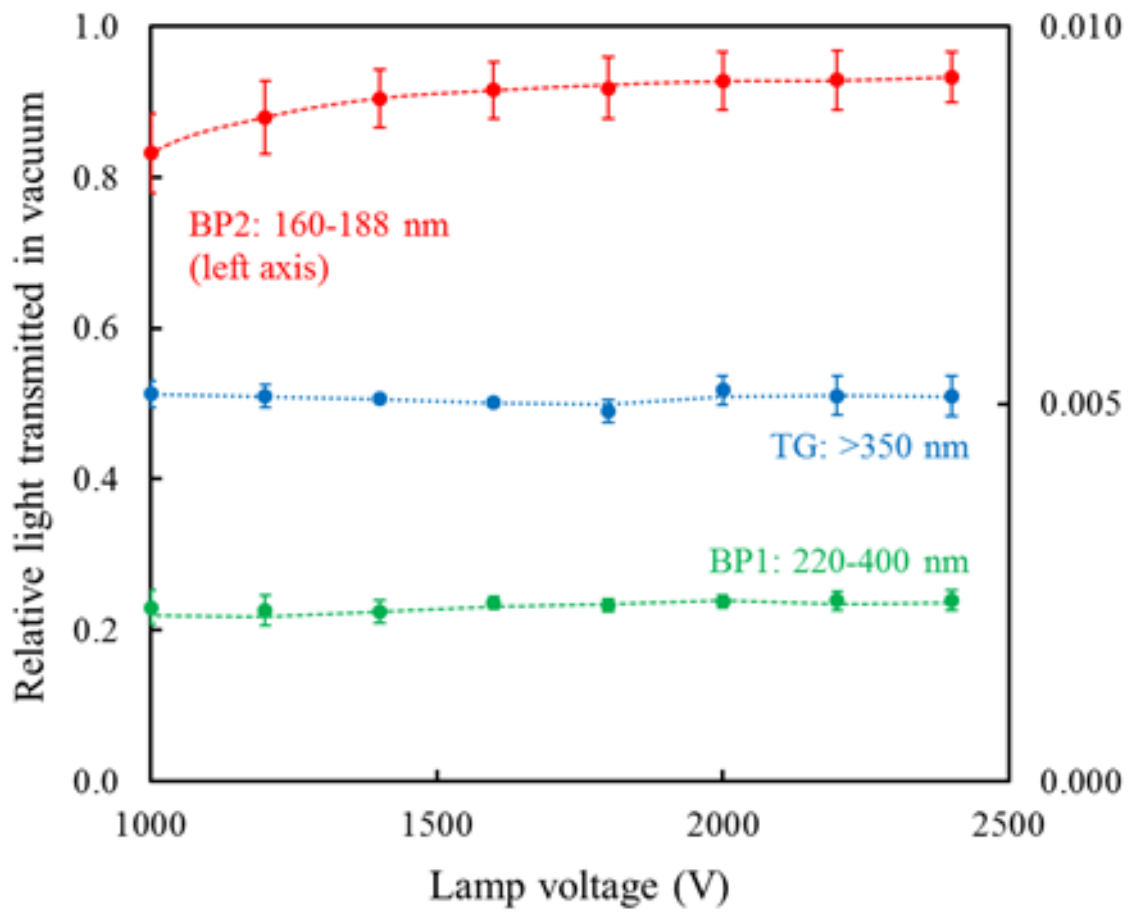

Fig. 3. Ratio between the light collected at the PMT with and without filter, as a function of the voltage applied to the lamp anode, measured in vacuum for different filters. The series in red (160-188 nm) reports to the left axis, while the series in green (220-400 nm) and blue (>350 nm) report to the right axis. The values were corrected for each filter transmission.

## 3. Results

In order to detect and measure the eventual reemission of Xe scintillation by TMA molecules, the chamber was filled with Xe-TMA mixtures and the light, coming directly from the lamp or reemitted, was collected at the PMT. The initial study was made without any filter. In Fig. 4 the MCA spectra obtained for the different Xe-TMA mixtures are presented. We can observe a gradual decrease of the centroid channel as TMA concentration increases, related with the reduction of the light that is collected by the PMT, due to absorption by TMA.

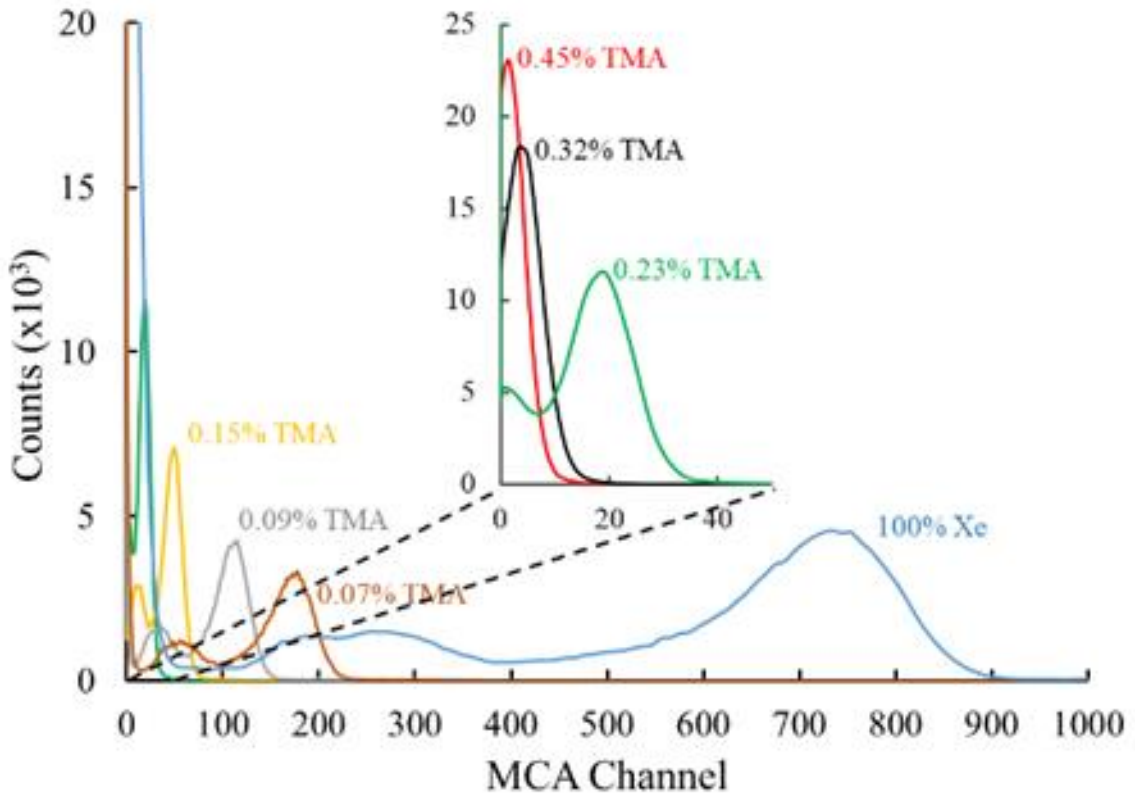

Fig. 4. MCA spectra of the light (160-650 nm) collected by the PMT for different Xe-TMA mixtures in the chamber. The total gas pressure is 800 Torr. The voltage applied to the lamp was 2000 V. A zoom is shown for the mixtures with 0.45%, 0.32% and 0.23% TMA fraction.

When possible, the MCA software fitted a Gaussian curve to the spectra and calculated the centroids of these distributions. When the light reaching the PMT is so low that the obtained spectra involved only a few channels near the origin, the MCA software was not capable of performing the fit. In these critical cases, the centroid positions were calculated through the weighted mean of the number of counts in each channel, as explained below. Above 0.5% of TMA at 800 Torr, the photons reaching the PMT were not enough to produce a visible signal and only residual light was detected.

To clarify, for each Xe-TMA mixture, if the light collected was coming directly from the lamp (160-188 nm) or if it was being reemitted by TMA (260-350 nm), the measurement was repeated for every mixture using the 220-400 nm BP1 bandpass filter. The corresponding spectra are presented in Fig. 5. As can be seen, with this filter, the fraction of light collected at the PMT was always the same, either in vacuum, in pure xenon or in the different mixtures, with the mean centroid position at channel 1.64±0.12 corresponding to about 0.2% of the total amount of light emitted by the lamp that reaches the PMT without filter and in vacuum (channel 708 in the MCA).

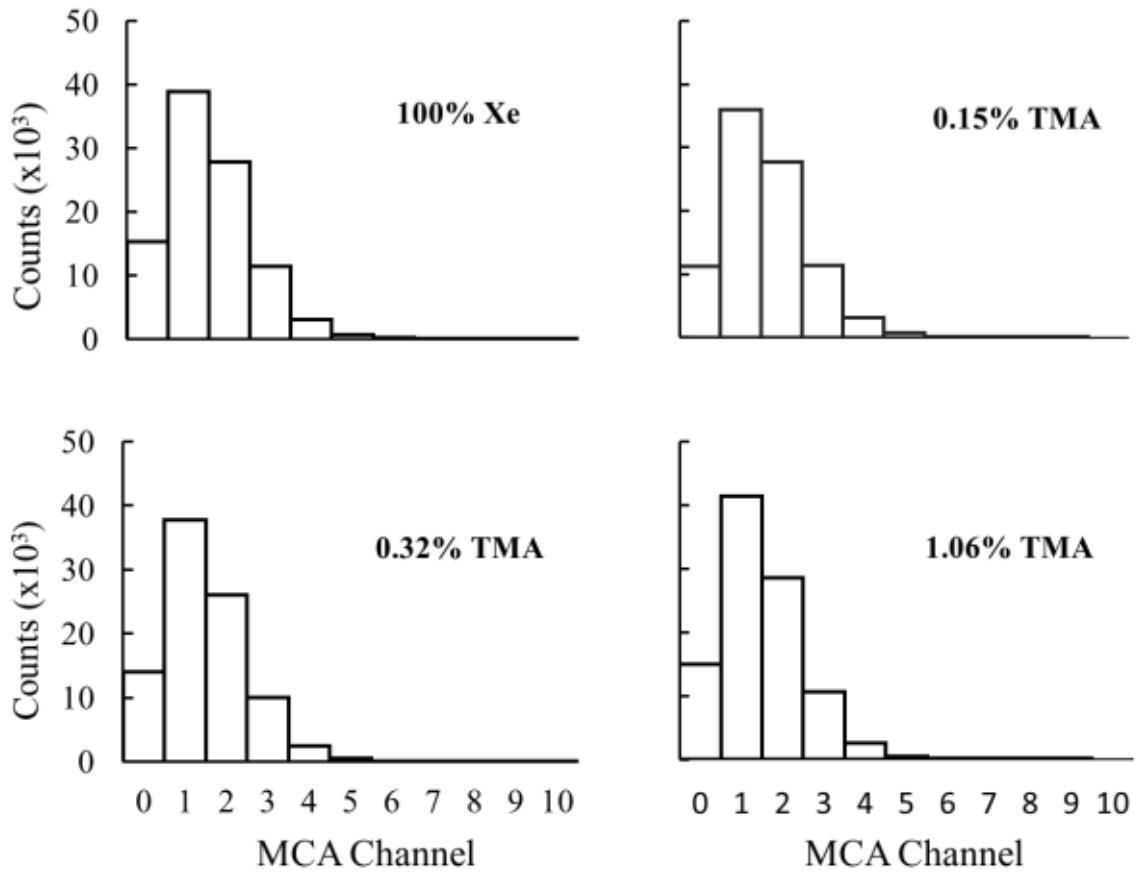

Fig. 5. MCA histogram of the light collected by the PMT using the bandpass filter BP1 (220-400 nm) for different Xe-TMA mixtures in the chamber. The total gas pressure is 800 Torr. The voltage applied to the lamp was 2000 V.

The results are summarized in Fig. 6, where the centroid channel of the MCA spectra is depicted as a function of TMA concentration in the mixtures. It can be seen that, without filter, the light collected decreases progressively as TMA percentage increases up to a partial TMA pressure of about 4 Torr (0.5% TMA at 800 Torr) (green symbols, left axis). With the bandpass filter of 220-400 nm on the PMT, the collected light is always the same and centered at channel 1.64 (red symbols, right axis). In Table 2 we summarize the results obtained for the light detected with the different bandpass filters used, relative to the light in vacuum without filter, in Xe and Xe-TMA mixtures at 800 Torr total pressure.

It can thus be concluded that not only TMA absorbs Xe scintillation, even at low percentages (above 4 Torr - 0.5% concentration - in an estimated average path of 50.3 mm), but also that it does not reemit it in the 260-350 nm range as far as our system could detect.

The absorption coefficient of TMA in the 160-188 nm range was calculated from the slope of the linear fit to the logarithmic experimental values without filter (Fig. 6). The value of 0.43±0.03 $Torr^{-1}cm^{-1}$ was obtained, in agreement with previous measurements performed at low pressure [19].

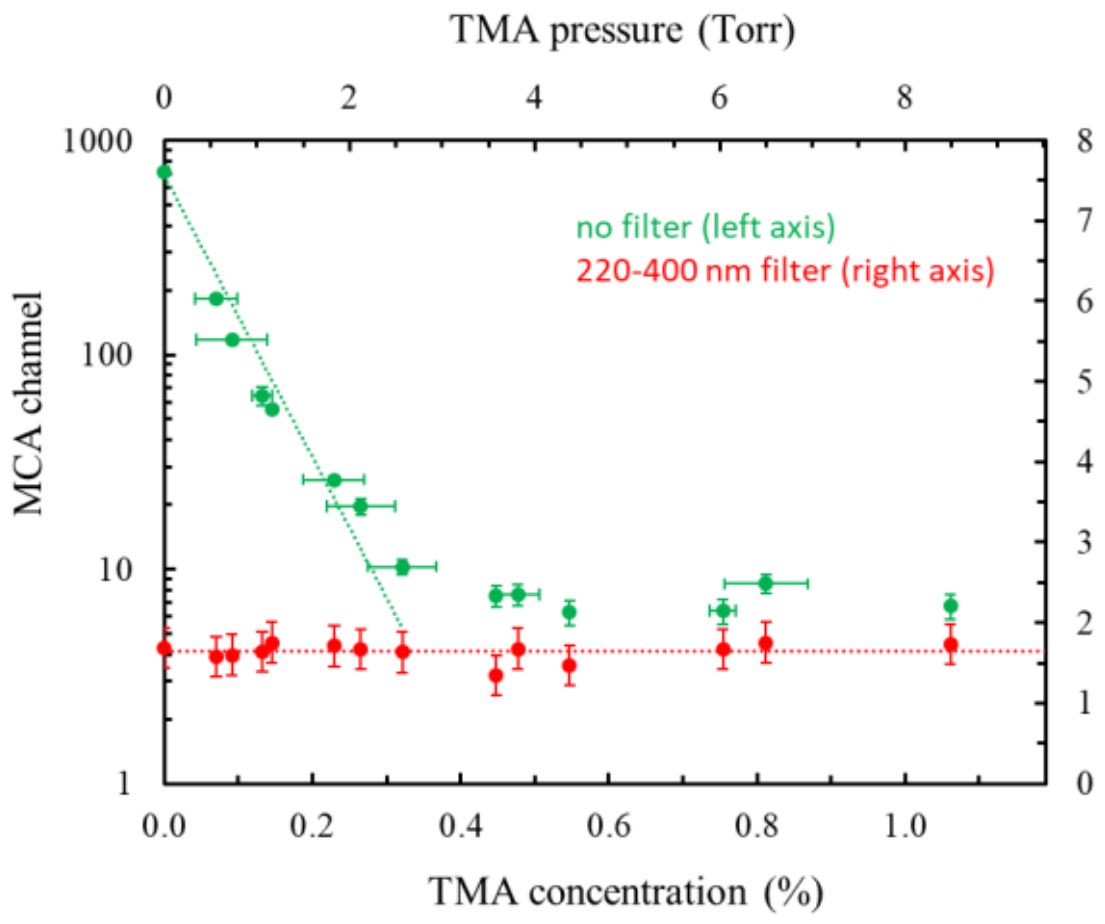

Fig. 6. Light collected at the PMT (MCA channel) as a function of TMA concentration in the mixture, at 800 Torr in an estimated average path of 50.3 mm, without filter (green symbols, left axis) and with the 220-400 nm filter (red symbols, right axis). The voltage applied to the lamp was the same in all cases (2000 V). Horizontal error bars are the same in both series. Vertical error bars in the data without filter, although represented, are small and not visible.

Table 2

Light detected (relative to the light in vacuum without filter) in pure Xe and Xe-TMA mixtures at 800 Torr total pressure, without filters and for the different bandpass filters used (at 2000 V lamp voltage).

| TMA (Torr) | TMA (% at 800 Torr) | No filter (%) | 220-400 nm (%) | >350 nm (%) | 160-188 nm (%) |
|---|---|---|---|---|---|
| 8.49 ± 0.45 | 1.06 ± 0.06 | 0.94 ± 0.07 | 0.24 ± 0.05 | 0.50 ± 0.05 | * |
| 6.03 ± 0.14 | 0.75 ± 0.02 | 0.89 ± 0.07 | 0.22 ± 0.05 | 0.45 ± 0.05 | * |
| 4.38 ± 0.23 | 0.55 ± 0.03 | 0.88 ± 0.07 | 0.22 ± 0.05 | 0.48 ± 0.05 | * |
| 3.58 ± 0.04 | 0.45 ± 0.01 | 1.05 ± 0.11 | 0.25 ± 0.05 | 0.48 ± 0.05 | * |
| 2.57 ± 0.37 | 0.32 ± 0.05 | 1.44 ± 0.07 | 0.23 ± 0.05 | 0.47 ± 0.05 | 0.7 ± 0.5 |
| 1.83 ± 0.33 | 0.23 ± 0.04 | 3.65 ± 0.07 | 0.24 ± 0.05 | 0.46 ± 0.05 | 1.6 ± 1.0 |
| 1.16 ± 0.04 | 0.15 ± 0.01 | 7.81 ± 0.10 | 0.19 ± 0.05 | 0.49 ± 0.05 | 6.6 ± 1.0 |
| 0.73 ± 0.38 | 0.09 ± 0.05 | 16.5 ± 0.2 | 0.21 ± 0.05 | 0.49 ± 0.05 | 11.2 ± 1.0 |
| 0.56 ± 0.23 | 0.07 ± 0.03 | 25.5 ± 0.9 | 0.23 ± 0.05 | 0.51 ± 0.05 | 24.5 ± 1.0 |
| 0.00 | 0.00 | 100 | 0.24 ± 0.05 | 0.49 ± 0.05 | 93.2 ± 4.4 |

* not measurable

## 4. Discussion

Although our experimental results indicate that there is no reemission of Xe light by TMA molecules, the upper reemission limit that can be established from these measurements depends on the detection limit of our experimental system. In order to assess this limit and to infer from it an upper reemission probability of Xe scintillation by TMA molecules, further analysis was made by a Monte Carlo simulation.

First of all, the limit of light detection of the MCA had to be estimated. In our best working experimental conditions and when the light collected in vacuum at the PMT with no filter (maximum light collected) was in channel 708 in the MCA, we considered that the smallest change that could be meaningfully detected in the critical region referred before (near the origin of the MCA) was 0.5 of a MCA channel. This corresponds to 0.07% of the maximum light collected at the PMT.

However, due to the isotropy of the eventual light reemission by TMA molecules, its detection efficiency by the PMT will be different from that due to the lamp. In fact, the reemitted photon besides having a higher wavelength can be emitted in all directions, and at different reflection coefficient in the chamber walls is higher than for VUV light. Figure 7 sketches the detection of the light without (a) and with reemission (b). The balance between the photons that can, through reemission either reach the PMT or miss it is important in the calculation of the reemission probability from the experimental data. To estimate the result of this balance, a Monte Carlo simulation model was developed. In this model all the characteristics and geometric details of the experimental setup were included, as well as the optical effects capable of changing the photons' direction, such as the refraction in the transmission windows and specular reflection of the different energy photons in the inner polished walls of the device, chosen according to the different wavelengths involved (VUV and ~300nm).

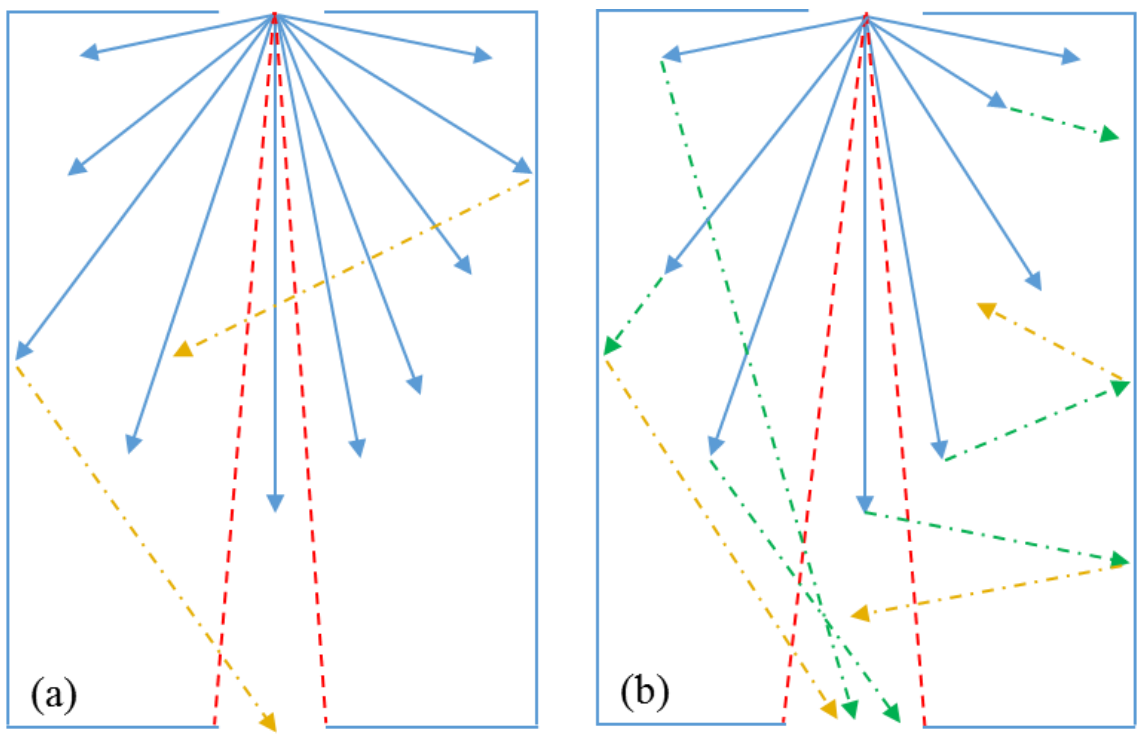

Fig. 7. Schematic representation of the light detection (a) without reemission and (b) with reemission. Blue lines represent Xe lamp photons and the red dashed lines the solid angle subtended by the PMT. Green dashed lines represent the reemitted photons and orange dashed lines photons reflected on the walls.

The flowchart of the Monte Carlo simulation is presented in Fig. 8. Each simulation run (for vacuum and for all the mixtures) considered $10^8$ photons entering the chamber coming from isotropic emission from the lamp. The initial photons' wavelength was chosen from a Gaussian distribution centered at 172 nm with 14 nm FWHM [28], reproducing the xenon VUV emission. The change in this distribution and in the photon's direction due to the window transmission and refraction, respectively, was also considered.

The simulation uses the TMA VUV photon absorption coefficient obtained from our experimental results and includes the possibility of reemission following photon absorption by TMA, considering different possible re-emission probabilities. The reemitted photon wavelength was chosen from the wavelength distribution in [23].

Each photon is followed in its path in the chamber, suffering eventually reflection on the inside surfaces of the chamber – either suprasil® or stainless steel – until it is absorbed by TMA (only for VUV photons), by the surfaces, or transmitted through the windows. The simulation checks if the photons transmitted through the exit window are detected by the PMT and if so they are counted as reemitted or as coming from the lamp, depending on their wavelength. The photon transmission and reflection coefficients, the refraction index of the suprasil® window and the photomultiplier quantum efficiency were taken from their data sheet [24, 25]. The reflection coefficient for the chamber stainless steel was obtained from [29] for wavelengths above 250 nm and from [30] for the VUV.

To relate experimental and simulation outputs, the simulation results in vacuum were normalized to the experimental ones, in vacuum (channel 708 of the MCA). Using this relation, the number of simulated photons that reach the PMT was converted into a channel number.

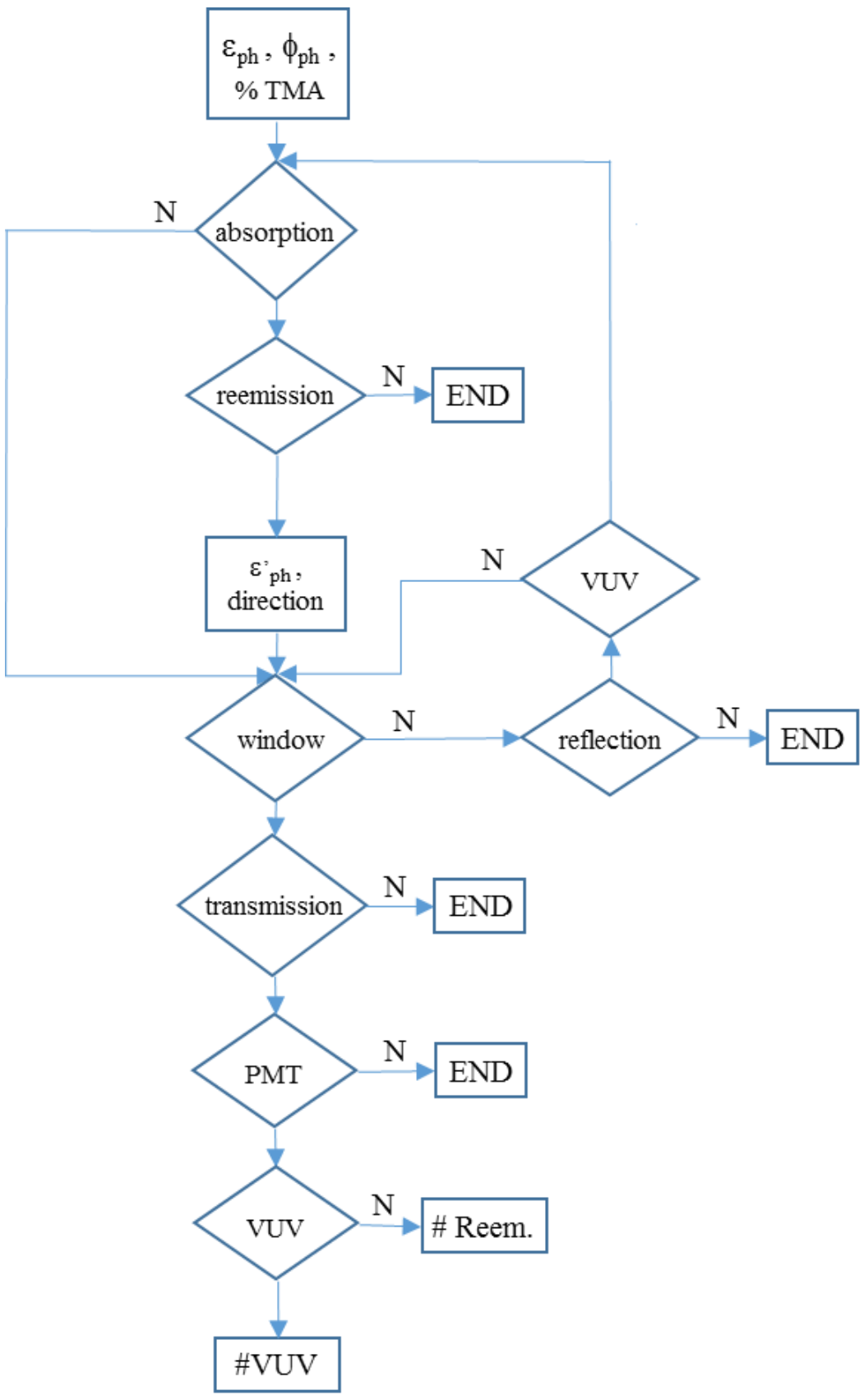

Fig. 8. Flow chart of the simulation carried out to assess the reemission probability of the photons eventually reemitted by TMA.

For each Xe-TMA mixture, the simulation was run considering different reemission probabilities, starting arbitrarily with the reemission of 0.1% of the absorbed photons, counting the reemitted photons that reach the PMT and converting them into a MCA channel. These results are presented in Fig. 9 where each curve represents the expected channel in the MCA for a given reemission probability, as a function of the TMA percentage. The nearly constant MCA channel value for TMA concentration above 0.2%, indicates that almost all lamp photons are absorbed in the mixture (~97%), and from then on, the number of reemitted photons is approximately constant, depending mainly on the reemission probability used.

The horizontal full line represents the estimated detection limit of 0.5 channel of the MCA. The reemission probability corresponding to this limit is the lowest reemission probability that would be detectable in our experimental conditions. Since, within our experimental conditions, no reemission was observed these results allow us to conclude that if reemission does occur, it must be below 0.3% of the light absorbed by TMA, otherwise it would have been experimentally observed.

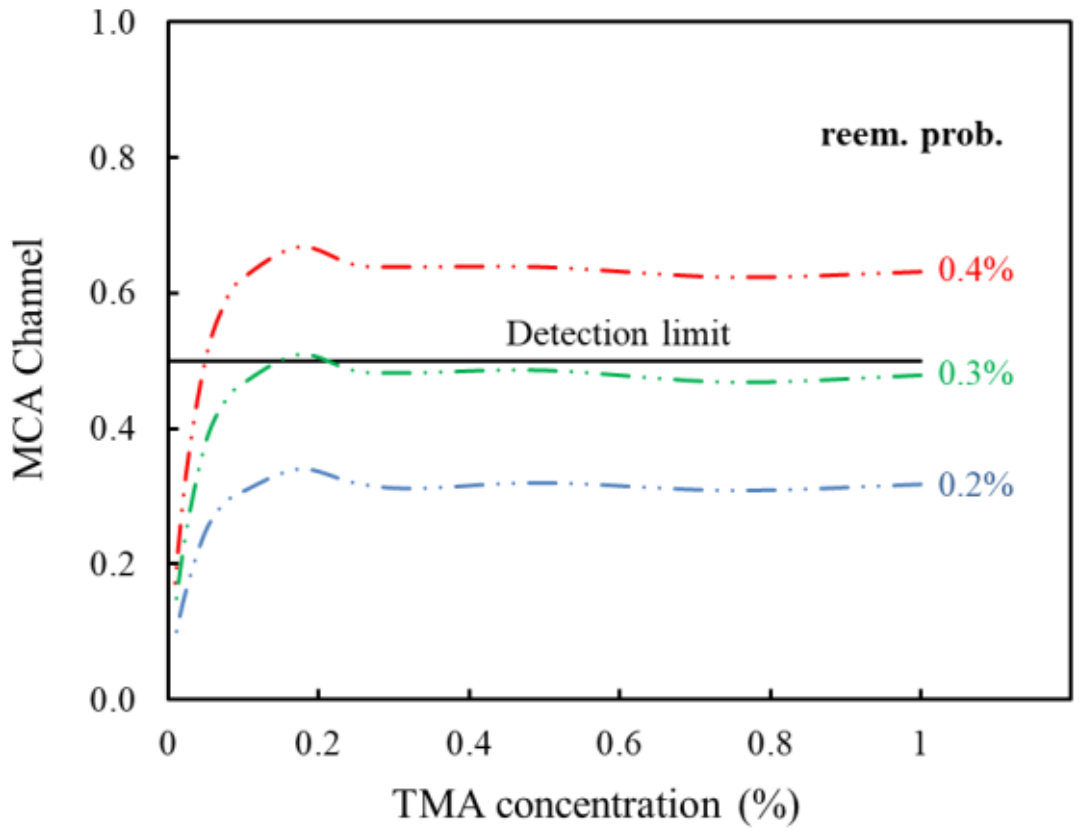

Fig. 9. Monte Carlo simulation results for number of reemitted photons (converted to channel in the MCA) as a function of TMA concentration, for four different reemission probabilities.

## 5. Conclusions

In order to assess the behavior of TMA regarding the reemission of xenon scintillation in the known TMA emission range of 260-350 nm, a special experimental setup was built. It consisted of a cylindrical stainless steel chamber 49.7 mm long, with two opposing suprasil® quartz windows in contact, one with a xenon lamp that emitted photons into the chamber and another with a photomultiplier tube that detected light that, after traversing the gas, was not absorbed by the gas molecules or that was reemitted. The light collected in the photomultiplier tube created a peak in a multichannel analyzer, whose centroid position was proportional to the intensity of the light detected. To quantify the wavelength distribution of the light emitted from the lamp and to identify the TMA absorption and expected reemission, adequate bandpass filters were used in vacuum, in pure xenon and in Xe-TMA mixtures with compositions in the range 0.07% - 1.06% TMA. All mixtures considered had their composition checked with a residual gas analyzer. In all experiments, the total gas pressure was 800 Torr and the gas was continuously purified with suitable getters. For TMA percentages above 0.5% in an estimated average path of 50.3 mm, xenon VUV light is absorbed and no signal can be observed in the MCA. An absorption coefficient of 0.43±0.03 $Torr^{-1}cm^{-1}$ was estimated for TMA in the 160-188 nm range, a value in agreement with previous measurements. Concerning TMA reemission in the 260-350 nm range, it was not observed within our estimated experimental precision (0.07% of the total light collected), for any TMA concentration. To take into account the isotropy of the reemission process and estimate an upper limit of detectable reemission in our experimental system, a Monte Carlo simulation was implemented. The Monte Carlo model used the measured experimental TMA absorption coefficient and included the relevant geometrical and optical details of the experimental system, including a wavelength dependent reflection of the radiation in the chamber's inner walls. Normalizing experimental and simulation results in vacuum conditions, reemission probability values could be scanned in order to reproduce the 220-400 nm filter experimental results in the mixtures, until the experimental detection limit was achieved. The simulation results have shown that a reemission probability higher than 0.3% should be detectable in our experimental system. Since no detectable reemission was observed, we concluded that if reemission occurs, its probability is below 0.3%.

Concerning the use of TMA as a dopant in a high-pressure TPC searching for $0\nu\beta\beta$, and in particular the question of the possibility to detect a robust, primary scintillation, $t_0$ signal, we believe that it will be quite challenging. In fact, the excited Xe dimer that follows the initial interaction has two radiative decay channels: a 172 nm direct radiative emission that will be strongly absorbed by TMA molecules without reemission (or <0.3%) as proven in this work; or, with a low probability (~3%, [18]), through fluorescent transfer to TMA, with subsequent emission of ~300 nm radiation. In either case, it appears unrealistic that a primary scintillation signal, even at the $Q_{\beta\beta}$ energy, will be detectable.

## 6. Acknowledgments

The NEXT Collaboration acknowledges support from the following agencies and institutions: the European Research Council (ERC) under the Advanced Grant 339787-NEXT; the Ministerio de Economía y Competitividad of Spain under grants FIS2014-53371-C04 and the Severo Ochoa Program SEV-2014-0398; the GVA of Spain under grant PROMETEO/2016/120; the Portuguese FCT – Fundação para a Ciência e Tecnologia - through the project PTDC/FIS-NUC2525/2014; the U.S. Department of Energy under contracts number DE-AC02-07CH11359 (Fermi National Accelerator Laboratory) and DE-FG02-13ER42020 (Texas A&M); and the University of Texas at Arlington. Alexandre M.F. Trindade was supported by FCT – Fundação para a Ciência e Tecnologia (SFRH/BD/116825/2016). José Escada was supported by FCT – Fundação para a Ciência e Tecnologia (SFRH/BPD/90283/2012). André F.V. Cortez received a PhD schorlarship from FCT – Fundação para a Ciência e Tecnologia (SFRH/BD/52333/2013).